# Exploring the origins of the Dzyalloshinski-Moriya interaction in MnSi

C. Dhital[1]*, L. DeBeer-Schmitt[2], Q. Zhang[1, 2], W. Xie[3], D. P. Young[1], J. F. DiTusa[1]**

[1] *Department of Physics and Astronomy, Louisiana State University, Baton Rouge, LA 70803*

[2] *Oak Ridge National Laboratory, Oak Ridge, Tennessee 37831, USA*

[3] *Department of Chemistry, Louisiana State University, Baton Rouge, LA, 70803, USA*

**Abstract:**

By using magnetization and small-angle neutron scattering (SANS) measurements, we have investigated the magnetic behavior of the $Mn_{1-x}Ir_xSi$ system to explore the effect of increased carrier density and spin-orbit interaction on the magnetic properties of MnSi. We determine estimates of the spin wave stiffness and the Dzyalloshinski-Moriya (DM) interaction strength and compare with $Mn_{1-x}Co_xSi$ and $Mn_{1-x}Fe_xSi$. Despite the large differences in atomic mass and size of the substituted elements, $Mn_{1-x}Co_xSi$ and $Mn_{1-x}Ir_xSi$ show nearly identical variations in their magnetic properties with substitution. We find a systematic dependence of the transition temperature, the ordered moment, the helix period, and the DM interaction strength with electron count for $Mn_{1-x}Ir_xSi$, $Mn_{1-x}Co_xSi$, and $Mn_{1-x}Fe_xSi$, indicating that the magnetic behavior is primarily dependent upon the additional carrier density, rather than on the mass or size of the substituting species. This indicates that the variation in magnetic properties, including the DM interaction strength, is primarily controlled by the electronic structure, as Co and Ir are isovalent. Our work suggests that although the rigid band model of electronic structure, along with Moriya's model of weak itinerant magnetism, describes this system surprisingly well, phenomenological models for the DM interaction strength are not adequate to describe this system.

**Introduction:**

The nanoscale twisted spin textures known as magnetic skyrmion lattices are of considerable interest among condensed matter physicists and material scientists, owing to the fundamental interactions generating such unusual textures and the potential for application in spintronic devices [1-5]. After the discovery of this magnetic structure in MnSi in 2009 [1], a similar structure was uncovered in several other non-centrosymmetric magnetic compounds [6-11]. All of these materials have a qualitatively similar phase diagram consisting of paramagnetic, helical, conical, and skyrmion lattice (A-phase) phases [1]. A prerequisite for the formation of these spin textures is the formation of a helical magnetic state with definite chirality.

Generally, the interplay between the antisymmetric DM interaction ($D$) and the uniform exchange interaction ($J$) produces a helical structure with a small wave vector $k \sim D/J$. However, the origin and size of $D$ in itinerant magnets has been the subject of recent theoretical activity, and understanding how to control its magnitude will be key for future materials design [12-15]. Its importance is also reflected in that the combination of the sign of $D$ and the chirality of the crystal lattice determine the chirality of the helix [16]. Other details of the helical state are determined by better-understood parameters, such as the smaller scale anisotropic exchange interaction (*AEI*), which controls the propagation direction for the helix, and the weak cubic anisotropy, which determines the spin wave gap, as well as some specifics of the magnetic structure under the application of magnetic field. The transition temperature $T_C$, $k$, handedness of the helix, and the propagation direction vary among different compounds depending upon the relative importance of each of these interactions [16-19].

These interactions, and hence the electronic and magnetic properties, can largely be controlled either by chemical substitution or by application of hydrostatic/uniaxial pressure [20-24]. MnSi has been extensively studied under different physical environments and with different chemical substitutions to probe the effects on the magnetic structure. In fact, controlled chemical substitution provides an opportunity to tune the fundamental interactions that are strongly coupled to the details of the electronic structure, the crystal symmetry, and the strength of the spin-orbit interaction. Since the size of the skyrmions, and hence the skyrmion density, depends upon two interactions, $D$ and $J$, it is also of practical importance to be able to control these parameters. Previous studies of chemically doped systems have shown that the transition temperature $T_C$, the ordered moment $M_S$, and the helix period $\lambda$ ($\lambda=2\pi/k$) are strong functions of the transition metal constituent and the level of substitution [16,21-22,25-28]. Nonetheless, predicting the effect on the magnitude and sign of $D$ due to chemical substitution or pressure remains largely elusive. Models of insulating magnets emphasize the degree of inversion-symmetry breaking evident in the crystal lattice and the size of the spin-orbit coupling constant [29,30]. However, these models cannot account for the large variation in helical periods, the handedness of the chirality, the magnitude of the coercive field found in the transition metal monosilicides and germanides, and the substitution series connecting them, all having the *B20* crystal structure with similar lattice constants and structural parameters [16,26,31-32]. More recently, models based upon the details of the electronic structure in proximity to the Fermi level, specifically anticrossing points, have had some success in describing the broad features of one substitution series, Mn$_{1-x}$Fe$_x$Ge [12-13]. To explore further the dependence of the important interaction energies on the spin-orbit coupling parameter and the electronic structure in this class of compounds, we investigated Ir substitution for Mn in MnSi, Mn$_{1-x}$Ir$_x$Si. Surprisingly, our data are almost identical to that of previous investigations of Mn$_{1-x}$Co$_x$Si and Mn$_{1-x}$Fe$_x$Si [10,16,28], emphasizing the importance of electronic structure for determining both $J$ and $D$.

The following sections summarize the results of magnetization and small angle neutron scattering measurements of the as-of-yet unexplored system Mn$_{1-x}$Ir$_x$Si, where a much heavier element Ir ($Z=77$) is substituted for Mn ($Z=25$). We discover a systematic decrease of $T_C$ and $M_S$ and a systematic increase of $k$ as function of $x$.

**Experimental Details:**

Single crystals of Mn$_{1-x}$Ir$_x$Si ($x < 0.06$) were synthesized by loading arc melted polycrystalline pellets made up of ultra-pure elemental constituents (>99.99% pure) inside graphite tubes and employing a modified Bridgman method in a RF furnace under a flowing argon environment. Attempts to synthesize phase pure single crystals for higher Ir concentrations at ambient pressure were unsuccessful, indicating the solubility limit for this substitution. The phase purity, crystallinity, and the stoichiometry of the samples were determined using powder X-ray diffraction, single crystal X-ray diffraction, and Wavelength Dispersive Spectroscopy (WDS) techniques. The details of sample preparation and the variation of lattice parameter with $x$ are presented in the supplementary materials (Fig. S1)[33]. Magnetization measurements, both ac and dc, were carried out in a Quantum Design 7-T MPMS SQUID magnetometer. The ac susceptibility measurements were performed at a frequency of 100 Hz with an ac driving amplitude of 1 Oe. Small Angle Neutron Scattering (SANS) measurements were carried out at the GP-SANS beamline at the High Flux Isotope Reactor (HFIR) at Oak Ridge National Laboratory (ORNL). All of the crystals were aligned such that the [1 -1 0] crystal direction was along the magnetic field which was oriented parallel to incident beam. In addition, the crystalline [1 1 1] direction was oriented such that it was nearly horizontal. The mean wavelength of incident neutrons employed was $\lambda=4.75$ Å with $\Delta\lambda/\lambda=0.16$ with a sample-to-detector distance of 8.65 m.

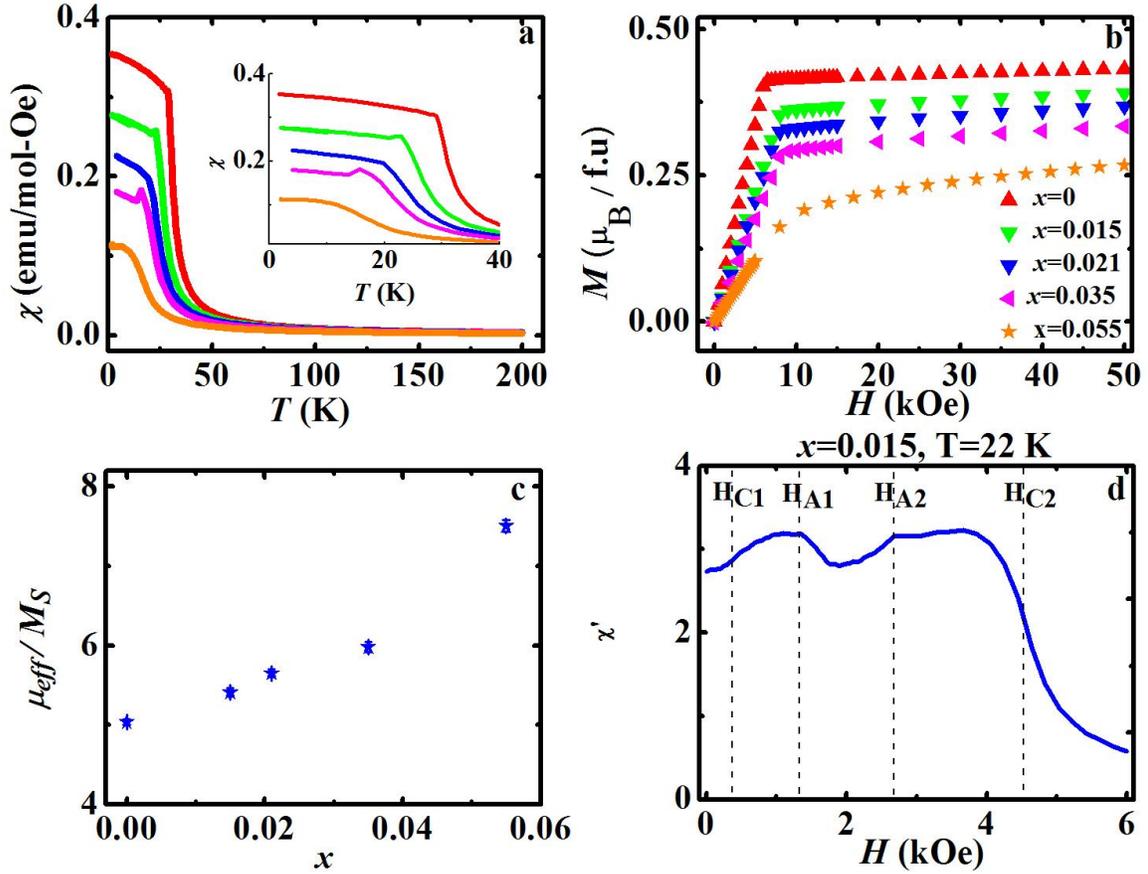

**Fig. 1** Magnetic properties of Mn$_{1-x}$Ir$_x$Si. (a) dc susceptibility, $\chi$, as a function of temperature, $T$. Line colors are the same as symbol colors identified in the key of frame b. (b) Magnetization, $M$, as a function of magnetic field, $H$, at 4 K (c) Rhodes-Wolfarth Ratio ($\mu_{eff}/M_S$) versus concentration $x$. Here, $\mu_{eff}$ is the effective moment obtained by fitting the modified Curie-Weiss form to the high temperature susceptibility, and $M_S$ is the saturated ordered moment at 4 K. (d) Real part of the ac susceptibility, $\chi'$ for $x = 0.015$ at $T=22$ K.

## Magnetic property measurements

The results of the dc magnetization measurements are summarized in Figs. 1a, b and c. It is clear that the magnetic transition temperature, $T_C$, (Fig. 1a) and the ordered moment at low temperature (Fig. 1b) decrease monotonically with increasing $x$, similar to observations in Mn$_{1-x}$Co$_x$Si and Mn$_{1-x}$Fe$_x$Si [10,16,21,28]. For all samples, the high temperature paramagnetic susceptibility can be well fit with a modified Curie-Weiss law, $\chi = \chi_0 + \frac{C}{T-\theta}$, where $\chi_0$ is a temperature-independent background, $C$ is the Curie constant, and $\theta$ is the Curie-Weiss temperature. Similar to MnSi, $\theta$ is nearly equal to $T_C$, whereas the effective moment ($\mu_{eff}$) obtained from the Curie constant is significantly higher than the saturated ordered moment ($M_S$) at low temperature. Fig. 1c shows the variation of the Rhodes-Wohlfarth ratio (RW) defined by $RW = \frac{\mu_{eff}}{M_S}$ with $x$. The increase in value of the RW ratio with increasing $x$ indicates a progression toward weaker itinerant behavior [34, 35].

We have also measured the ac susceptibility as a function of dc field for several temperatures near $T_C$ for each of our crystals (see Fig. 1d and Fig. S2 [33]). A typical variation of the real part of the ac susceptibility with dc field is shown in Fig. 1d, where four characteristic fields $H_{C1}$, $H_{A1}$, $H_{A2}$, and $H_{C2}$ are indicated [4,10,36,37]. These transitions correspond to: the alignment of the magnetic domains, such that $k$ is along the field ($H_{C1}$, represented by the rapid increase in susceptibility at low field); the single magnetic domain state (referred to as conical) to the A-phase ($H_{A1}$, represented by the starting point of decreasing susceptibility); the A-phase back to the conical phase ($H_{A2}$, represented by the completion of the decreased susceptibility pocket); and conical phase to the field polarized phase ($H_{C2}$, represented by rapid decrease of susceptibility).

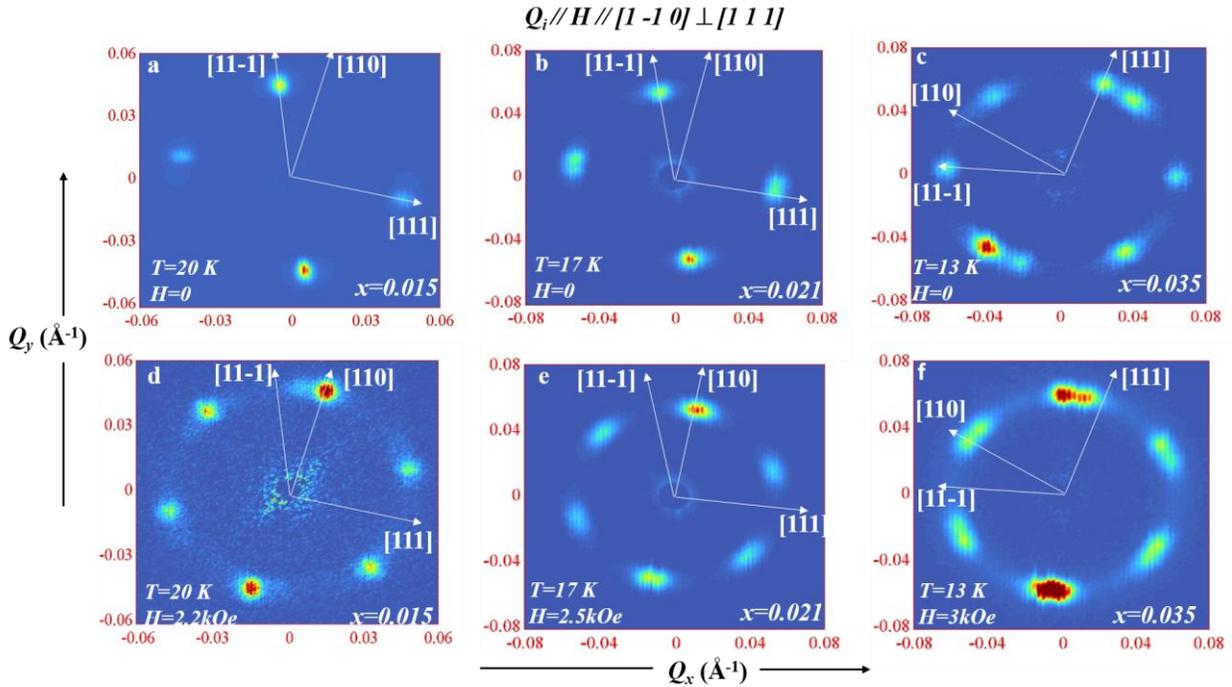

**Fig. 2** Small Angle Neutron Scattering (SANS) measurement on $Mn_{1-x}Ir_xSi$. The magnetic field was applied along the [1-10] direction parallel to the incident beam. White arrows represent the [1 1 1], [1 1 -1], and [110] directions in the plane perpendicular to the beam. (a) (b) and (c) display the scattering pattern in the helical phase, whereas (d), (e), and (f) represent the scattering pattern in the A-phase for $x = 0.015$, $x = 0.021$, and $x = 0.035$, respectively.

Interestingly, we did not observe such features in the ac susceptibility of our $x = 0.055$ crystal (Fig. S2e), which may indicate an absence of the A-phase at this level of chemical substitution [33].

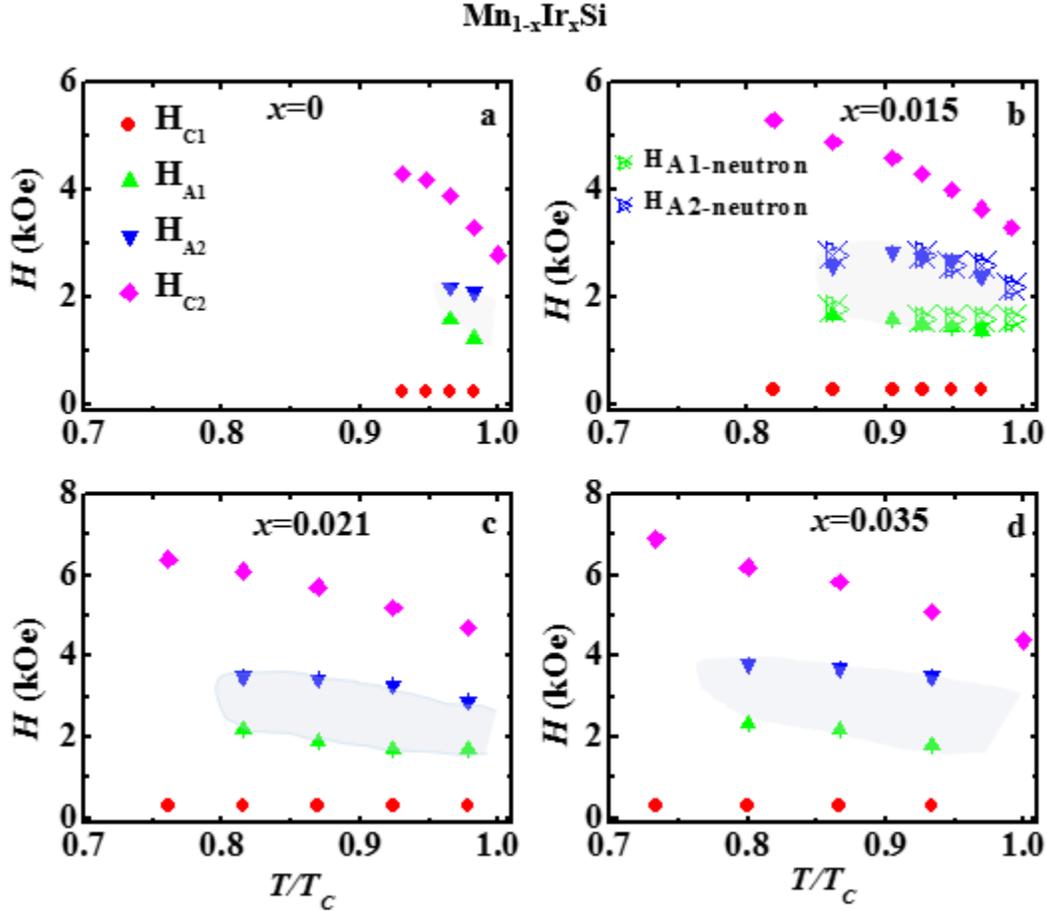

**Fig. 3**: Magnetic phase diagram of Mn$_{1-x}$Ir$_x$Si (a) for $x = 0$, (b) for $x = 0.015$, (c) for $x = 0.021$, and (d) for $x = 0.035$ determined from ac susceptibility measurements. The field values obtained from small angle neutron scattering measurements on the $x = 0.015$ crystal are plotted in (b). The shaded region represents the A-phase.

**Small Angle Neutron Scattering**:

Small angle neutron scattering measurements are ideal for exploring extended magnetic structures, such as the long period helical and A-phase states in MnSi. Typical scattering patterns that correspond to these phases are presented in Fig. 2, with Fig. 2a, b, and c presenting the scattering in the helical state ($H=0$, $T<T_C$). In the present experimental configuration, two out of the four equivalent [111] directions lie in the detector plane. For a single crystallographic domain sample in this sample orientation, we expect to observe four peaks corresponding to the equivalent [111] directions in the crystal. One pair of peaks is 180º apart due to the scattering along the [111] direction, while the other pair, at an angle of 70.5º (109.5º) from the first pair, corresponds to scattering along the [11-1] direction. For some samples, we also observe weak higher-order peaks arising from multiple scattering that is not visible at the intensity scale used in Fig. 2. The $x = 0.035$ sample contains a second, misaligned crystallographic domain, so that a third set of peaks is visible in the detector plane originating from a magnetic domain associated with this second crystallographic domain. However, our conclusions are not affected by the presence of the second crystal domain, as the magnetic scattering from this domain is clearly distinguished from the contribution of the main crystalline domain (Fig. S3 [33])). Fig. 2d, e, and f present the scattering in a finite magnetic field for $x = 0.015, 0.021$, and 0.035, respectively. This hexagonal intensity pattern was traditionally called the A-phase [37,38] and later became known as the skyrmion lattice phase after work by Mühlbauer *et.al* [1]. Consistent with previous results [1,38], the peak positions of the hexagonal scattering pattern are rotated from that of the helix and aligned along the [110] direction. This feature is present in all of our samples investigated via SANS. We have also observed a shallow ring of scattering just above $T_C$ in all samples (See supplementary materials Fig S5), which is a signature of the precursor fluctuating helical phase, as was seen in nominally pure MnSi in previous work [39,40].

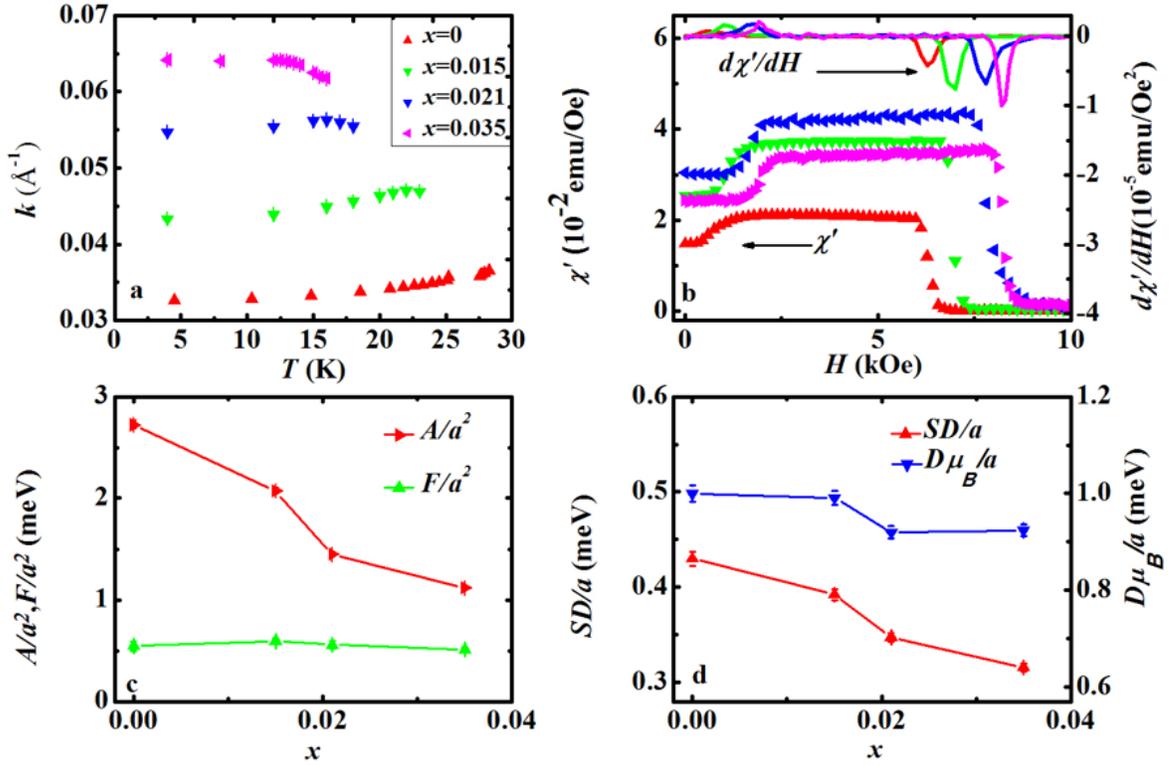

**Fig. 4** Parameterization of the magnetic states of $Mn_{1-x}Ir_xSi$. (a) Variation of the wave vector, $k$, in the helical phase as a function of temperature, $T$. Data for $x = 0$ are taken from Ref. [43] with permission of the publisher. (b) Real part of the ac susceptibility, $\chi'$, (left axis) and its field derivative, $d\chi'/dH$, (right axis) as a function of dc magnetic field at 4 K. The symbols and colors in frames a and b are the same. (c) Anisotropy constant, $F$, and spin wave stiffness, $A$, divided by the square of the lattice constant, $a^2$, as a function of $x$. (d) Dzyalloshinski-Moriya interaction, $SD/a$ (left), and $D/a=SD/M_Sa$ (right), where $D$ is the Dzyalloshinskii constant and $M_S$ is the ordered moment per Mn.

After confirming the presence of the helical and A-phases, we performed temperature and field scans for each of these samples. We were careful to control the field/temperature history prior to taking data, as each sample was heated to a temperature above $T_C$ and cooled to the desired temperature in zero field. Combining the results from ac susceptibility and SANS, we present magnetic phase diagrams for $Mn_{1-x}Ir_xSi$ in Fig. 3. The phase diagram is based mainly on results of the ac susceptibility measurements. The phase boundary for the A-phase of the $x = 0.015$ sample was also identified using SANS [Fig 3b]. For all other samples we verified the presence of the hexagonal scattering pattern at a few fields $H_{A1} < H < H_{A2}$. The variation of the integrated intensity of such patterns at different fields and temperatures is presented in Fig S6 [33]. These phase diagrams are qualitatively similar to that of nominally pure MnSi with only the field and temperature values modified. It is also evident that the relative region of stability of the A-phase increases as a function of $x$. Such an increased stability range has been observed previously in thin films, chemically doped systems, and in the presence of uniaxial/hydrostatic pressure [24,36,41,42]. It is not clear whether this is an electronic structure related change, or if it is due to disorder playing the same role as thermal fluctuations, as fluctuations are required for the formation of this phase in nominally pure MnSi. It is beyond the scope of this paper to explore the role of disorder on the fluctuating chiral phase above $T_C$, as was carried out for $Mn_{1-x}Fe_xSi$ [27], or if it is responsible for the increase in the stability range for the A-phase. A separate more detailed SANS experiment is required to answer these questions.

We have also traced the variation of $k$ in the helical state as a function of temperature and Ir concentration (Fig. 4a). It is clear that there is a significant increase in $k$ as a function of $x$. Similar to previous neutron diffraction studies [43], we also observe a slight decrease in $k$ upon cooling from $T_C$. Since in most treatments $D$ is expected to be temperature independent, the slight variation in $k$ with temperature is likely to be related to a slight modification in the ferromagnetic coupling due to spin fluctuations [43, 44]. However, the decreasing trend of $k$ with cooling is less obvious with increasing Ir concentration and its associated disorder, with our $x=0.021$ and $0.035$ samples showing an increase in $k$ with cooling near $T_C$.

In addition, we have characterized the critical behavior of our samples by fitting the variation in intensity of the magnetic scattering as a function of temperature by a standard mean-field power law model, $I = I_0(1-T/T_C)^{2\beta}$ for $T_C > T > 4$ K (See Fig. S4 [33]). For all samples, the value of the exponent $\beta \approx 0.25$, which indicates a tricritical mean field behavior as in nominally pure MnSi [45, 46], but which is distinct from the other magnetic *B20* materials. This is consistent with the previous work on MnSi that claimed that the magnetic transition in zero field is weakly first order due to critical fluctuations [46,47]. The difference from other *B20* materials, such as $Fe_{0.8}Co_{0.2}Si$ and FeGe, may be due to the relatively long range of the exchange interaction in MnSi and the presence of critical spin fluctuations, as pointed out in Ref. [46]. However, a recent study [48] gives evidence that the first order transition in zero field and the presence of precursor fluctuations are not related. Although our data are not sufficient to add to the discussion of the relationship of the spin fluctuations and the first order nature of the transition, we point out that the universality class does not change with Ir substitution.

**Estimation of Interaction parameters**

From the data presented above, we are able to determine several important magnetic parameters for each sample and present their dependence on $x$. Figs. 4b, c, and d summarize these parameters at 4 K. Fig. 4b presents representative ac susceptibility data that was used to determine $H_{C1}$ and $H_{C2}$ corresponding to the two peaks in the derivative with respect to $H$ of the ac susceptibility ($d\chi'/dH$). The determination of these fields allows us to estimate the spin wave stiffness $A$ ($A=g\mu_B H_{C2}/k^2$) and the anisotropy constant $F$ ($F=2g\mu_B H_{C1}/k^2$) [19, 27, 28]. The spin wave stiffness, $A$, is related to the magnetic field needed to destabilize the helical structure into the fully field-polarized state. The expression for $A$ is strictly valid for large momenta $Q \gg k$, i.e for distances smaller than the helical wavelength ($\lambda$), where the interaction between spins is essentially ferromagnetic. This approximation gives an estimate of the strength of the ferromagnetic exchange ($J$), which is proportional to $A$. The relation between $J$ and $A$ should be determined from inelastic neutron scattering measurements, as the analytic form is dependent on the model of magnetism used for analysis. Whether any of the common models is appropriate for MnSi is still an open question. $F$ determines the strength of the anisotropic exchange interaction and the cubic anisotropy. These expressions for $A$ and $F$ arise from an extension of the Bak-Jensen model [17], which takes into account the direction of the applied magnetic field with respect to the helix direction and the anisotropic interactions [27, 28]. The values of $A/a^2$ and $F/a^2$, where $a$ is the lattice constant, are plotted as a function of $x$ in Fig. 4c. It appears that there is no significant change in $F$ with $x$, whereas $A$ decreases significantly and monotonically with $x$. Although there is no unique universally accepted method to calculate $D$, one approach is to estimate the strength of the Dzyaloshinski-Moriya interaction ($SD$) and $D$ (Fig. 4d) by making use of the relation $SD = kA$, essentially connecting the helical wave vector to the ratio of $J$ and $D$ [19, 27]. Here, $S$ is the ordered moment per Mn atom. The right axis of Fig. 4d, $D/a$, is obtained by dividing $SD/a$ by the experimentally determined saturated magnetic moment per Mn atom, $M_S$.

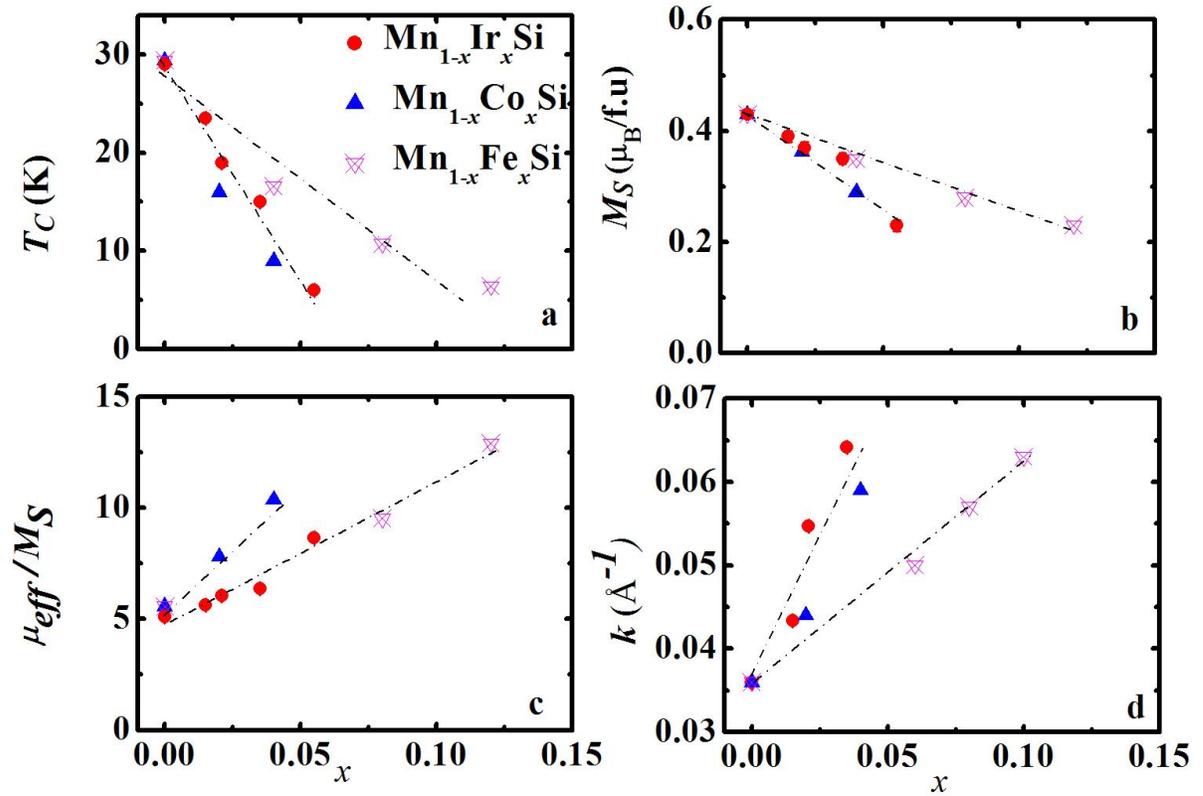

**Fig. 5** Comparison of magnetic properties of $Mn_{1-x}Ir_xSi$, $Mn_{1-x}Co_xSi$, and $Mn_{1-x}Fe_xSi$. Magnetic transition temperature, $T_C$ (a), ordered moment, $M_S$ (b), Rhodes-Wohlfarth ratio (c), and helix wave vector, $k$, at 4 K (d), as a function of $x$. Lines are a guide to the eye. The data for $Mn_{1-x}Co_xSi$ and $Mn_{1-x}Fe_xSi$ are reproduced from data in Refs. [10,27,49,50] with permission from the publishers and/or authors.

From Fig. 4d it is clear that *SD/a* decreases monotonically with $x$, mainly due to the decrease in *S*, whereas the Dzyaloshinskii constant surprisingly decreases slightly with $x$.

**Comparison with similar materials:**

To better understand the changes we observe with Ir substitution, we compare our data to the results of previous investigations of $Mn_{1-x}Co_xSi$ and $Mn_{1-x}Fe_xSi$ in Fig. 5 [10,27,49,50]. If we consider the cases of Ir, Co, and Fe substitution for Mn, three changes are expected. (i) A change in the carrier density due to the added valence electrons with substitution, which is two times as large for Ir and Co doping than for Fe. (ii) An increase in the spin-orbit interaction and hence *D* is expected from the relation: $\boldsymbol{D} = \zeta\, \boldsymbol{y} \times \boldsymbol{r_{12}}$, where $\zeta$ is the spin-orbit coupling strength that naively is expected to increase as $Z^4$, $y$ is a measure of the asymmetry of the crystal structure, and $r_{12}$ is the distance between interacting magnetic moments [29]. (iii) A slight change in the chemical pressure [36], which is positive for Fe and Co doping (decrease in unit cell volume) and negative for Ir doping (increase in unit cell volume). The comparison plots in Fig. 5 make clear that $Mn_{1-x}Ir_xSi$ and $Mn_{1-x}Co_xSi$ undergo nearly identical changes to $T_C$, $M_S$, and $k$ as a function of $x$. The variation of these parameters in $Mn_{1-x}Co_xSi$ has been previously shown to take place at twice the rate in $x$ as in $Mn_{1-x}Fe_xSi$ [10]. However, the variation of $\mu_{eff}/M_S$ is somewhat different in the Co and Ir doped systems indicating a slightly different trend in the degree of itinerancy. This suggests that the number of added valence electrons primarily controls the magnetic properties, whereas the change in spin-orbit interaction due to the larger mass of the Ir ions and the change to the lattice constant produce only secondary effects that are outside of our detection.

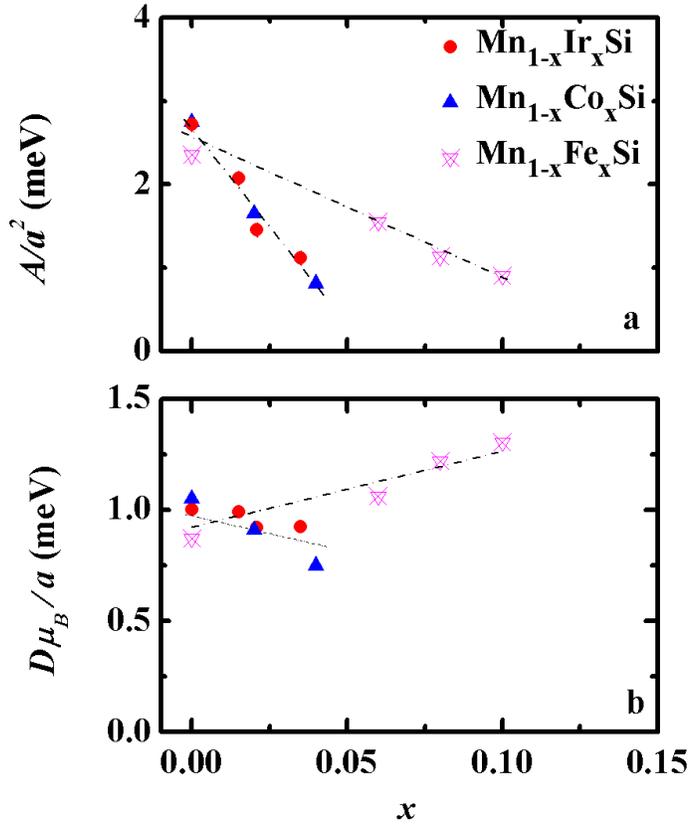

**Fig. 6** Comparison of interaction energies for Co, Mn, and Ir substitutions in MnSi. (a) $A/a^2$ and (b) $D\mu_B/a$ as a function of $x$ for $Mn_{1-x}Ir_xSi$, $Mn_{1-x}Co_xSi$, and $Mn_{1-x}Fe_xSi$ at low temperature. Here, $A$ is the spin wave stiffness parameter, $D$ is the Dzyaloshinskii parameter, and $a$ is the lattice constant. The data for $Mn_{1-x}Co_xSi$ and $Mn_{1-x}Fe_xSi$ are reproduced from references [10,27,49,50] with permission from the publisher and/or authors. Lines are a guide to the eye

We follow this comparison through to the interaction constants in Fig. 6 where $A$ and $D$ are presented for $Mn_{1-x}Ir_xSi$, $Mn_{1-x}Fe_xSi$, and $Mn_{1-x}Co_xSi$ at low temperature. Here, we have made use of values for the transition fields and the helix wave vectors reported in Refs 10, 27, 49, and 50. The variation of $A$ and $D$ as a function of $x$ is very similar for Co and Ir substitution, with $A$ changing similarly with Fe substitution at half the rate. This indicates that the variation of spin wave stiffness or the exchange interaction, $J$, where $J \propto A/S$ [19,27], is primarily controlled by the variation of the electronic structure which will vary systematically with electron count in a rigid band model. The variation of $D$ is not as simple to interpret with Fe substitution creating a moderate increase, while Co and Ir substitution results in a very similar slightly decreasing trend with $x$.

As mentioned earlier, there is no well-established method for estimating $D$. In our analysis presented in Figs. 4 and 6, we have relied on measurements of critical fields, $k$'s, and $M_s$'s along with the results of an extension to the model of Bak-Jensen [17,19], to make estimates of the important magnetic parameters $A$ and $D$. This model was specifically developed for the case of $B20$ materials and predicts values of spin wave stiffness $A$ for MnSi that are in good agreement with values found from inelastic neutron scattering [51]. However, when we make use of other methods for approximating these parameters, we find somewhat different values and trends. For example, assuming a finite temperature simple mean-field relationship between $T_C$ and $J$, $k_B T_C \approx JS^2$ [52] and that $D=kA/S$ with the standard assumption $A \approx 2JSa^2$ [52], the variation of $D\mu_B/a$ with substitution can be expected to vary as $2k_B T_C ka/S^2$. During our calculation $M_S$ replaces $S$. Following this method of estimation, the variation of $D/a$ among the silicide substitution series is shown in Fig. 7. This gives a significantly different dependence of $D/a$ as a function of $x$ when compared to Fig. 6b where we rely upon the critical field $H_{C2}$ to estimate $A$. We have used $H_{C2}$ determined at low temperature where mean field theory is expected to be a reasonable approximation. In contrast, the results of the analysis shown in Fig.7 depend upon the assumption that $A$ can be accurately determined from $T_C$. The differences evident in Figs. 6b and 7 may also indicate that $J$ or $A/M_S a^2$ may not be simply proportional to $T_C/M_S^2$, or that the relationship between $A$, $k$ and $H_{C2}$ is not straightforward.

In Fig. S7 [33] we plot $T_C$ as a function of $A/M_S a^2$ (with $A$ determined from the relation $A=g\mu_B H_{C2}/k^2$) for a large number of compounds that display the skyrmion lattice state [10,27,32,49,50,53-56]. Here, the general trend of an increasing $T_C$ with $A/M_S a^2$ is observed. However, a simple linear relationship is not well supported by the data, even when restricting consideration to MnSi derived materials.

In addition, to highlight the differences in estimates made via these two methods, we have presented a table of parameters for MnGe, $Mn_{1-x}Ir_xSi$, and FeGe in Table I. We find different values and trends for $A$, and $D$ in these three isostructural magnetic compounds. This confirms our conclusion that comparisons based upon simple mean field estimates, and the idea that $D$ is exclusively determined by the crystal symmetry, may not be reliable. Therefore, without more direct measurements of the interaction constants, estimates of $A$ and $D$ remain suspect, making a quantitative and convincing understanding of the origins and a reliable method for predicting the behavior of weak itinerant magnetism in non-centrosymmetric systems difficult.

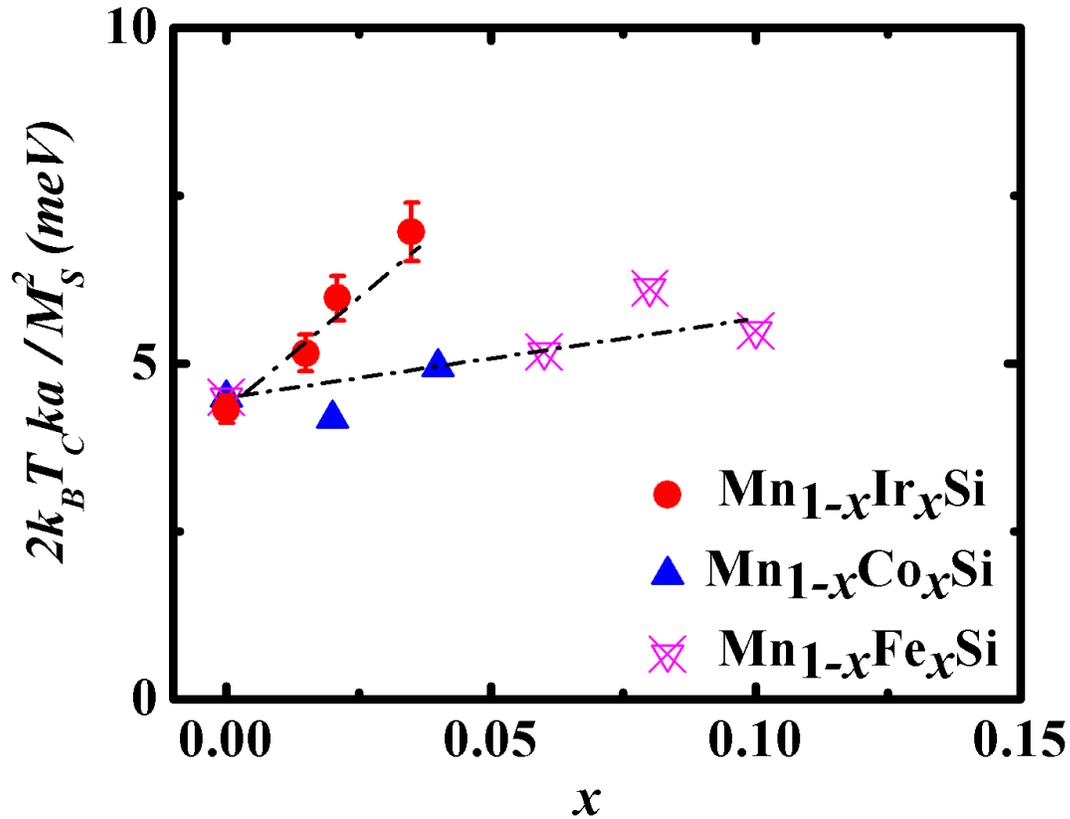

**Fig. 7** Comparison of the strength of the Dzyaloshinskii-Moriya interaction, $D/a$, as a function of substitution level, $x$, for MnSi, assuming $(D\mu_B)_{MF}/a \propto 2k_B T_C ka/M_S^2$. The data for $Mn_{1-x}Co_xSi$ and $Mn_{1-x}Fe_xSi$ are reproduced from references [10,27,49,50] with permission from the publisher and/or authors. Lines are a guide to the eye. The lattice parameter for $Mn_{1-x}Co_xSi$ and $Mn_{1-x}Fe_xSi$ are estimated using Vegard's law.

## Conclusion:

In conclusion, by exploring the magnetic behavior of $Mn_{1-x}Ir_xSi$ and comparing to other substitution series in MnSi, we have shown that the variation of the uniform exchange, the saturated magnetic moment, and the helical wave vector systematically vary with the change in the carrier density. The changes we measure in the magnetic properties are dominated by the variation of $A$ and $J$. These observations support the models for the substitution series in MnSi that make the simplifying assumption of a rigid band model of electronic structure and the Moriya theory of magnetism for this itinerant compound. Despite the expectation of increased spin-orbit coupling and the DM interaction with Ir substitution, we find no significant difference in the value of $D$ when compared to Co substitution. This further indicates that $D$ is determined primarily by the electronic structure, which is largely controlled by the electron density in the monosilicides and monogermanides [12]. A wider comparison of the *B20* compounds makes clear that mean field estimates that rely on $T_C$ to establish the important energy scales are insufficient for useful comparisons of $D$. We conclude that to make valuable comparisons to models of the magnetism in noncentrosymmetric magnets, experimental measurements of both $A$ and $k$ for a wide range of monosilicide and monogermanide transition metal compounds, and their substitution series, appears to be necessary. In the absence of these measurements, or a universally applicable method for determining $D$ more directly from measurement, a useful feedback between experiment and theory necessary for creating predictions of materials where the Dzyalloshinski-Moriya interaction is likely to produce novel and useful magnetic states will be lacking.

**Table I:** Comparison of parameters for MnGe, FeGe, and Mn$_{1-x}$Ir$_x$Si using measured values of the critical fields and results of the mean field model to make estimates. Data for MnGe, FeGe, and MnSi are reproduced from references [32,36,43,53,54] with permission from the publisher and/or authors.

| Compound | $T_C$ (K) | $S=M_S$ ($\mu_B$) | $k$ (Å$^{-1}$) | $H_{c2}$ (kOe) | $\frac{A}{a^2} = \frac{g\mu_B H_{c2}}{k^2 a^2}$ (meV) | $D/a = kA/M_S a$ (meV) | $(A/a^2)_{MF} \propto 2k_B T_C / M_S$ (meV) | $\frac{(D\mu_B)_{MF}}{a} \propto 2ka(k_B T_C)/M_S^2$ (meV) | Reference |
|---|---|---|---|---|---|---|---|---|---|
| MnGe | 170 | 1.9 | 0.21 | 128 | 1.46 | 0.77 | 15.42 | 8.18 | 32,53,54 |
| FeGe | 280 | 0.9 | 0.009 | 4 | 25.65 | 1.2 | 53.6 | 2.51 | 32 |
| MnSi | 29 | 0.42 | 0.035 | 6 | 2.72 | 1.05 | 11.90 | 4.52 | 36,43 |
| Mn$_{0.985}$Ir$_{0.015}$Si | 23 | 0.39 | 0.043 | 7 | 2.07 | 1.05 | 10.16 | 5.11 | This paper |
| Mn$_{0.979}$Ir$_{0.021}$Si | 19 | 0.37 | 0.055 | 7.8 | 1.45 | 0.99 | 8.85 | 5.89 | This paper |
| Mn$_{0.965}$Ir$_{0.035}$Si | 15 | 0.33 | 0.064 | 8.3 | 1.12 | 0.99 | 7.83 | 6.93 | This paper |
| Mn$_{0.945}$Ir$_{0.055}$Si | 6 | 0.23 | - | 8.7 | - | - | 4.49 | - | This paper |

**Acknowledgements**: The authors acknowledge Dr. Clayton Loehn and Dr. Nele Muttik at the Shared Instrumentation Facility (SIF) at LSU for assistance with the chemical analysis. We also thank D. A. Browne and R. Jin and I. Vekhter for helpful discussions. This material is based upon work supported by the U.S. Department of Energy under EPSCoR Grant No. DE-SC0012432 with additional support from the Louisiana Board of Regents. The SANS data and alignment of the samples used resources at the High Flux Isotope Reactor, a DOE Office of Science User Facility operated by the Oak Ridge National Laboratory. W. Xie was supported by the Louisiana Board of Regents Research Competitiveness Subprogram (RCS) under Contract Number LEQSF(2017-20)-RD-A-08 for the single crystal X-ray diffraction.

*cdhital@lsu.edu, **ditusa@phys.lsu.edu

## References:

1. Mühlbauer, S., Binz, B., Jonietz, F., Pfleiderer, C., Rosch, A., Neubauer, A., Georgii, R. and Böni, P., 2009. Skyrmion lattice in a chiral magnet. *Science*, *323*(5916), pp.915-919.

2. Fert, A., Cros, V. and Sampaio, J., 2013. Skyrmions on the track. *Nature nanotechnology*, *8*(3), pp.152-156.

3. Finocchio, G., Büttner, F., Tomasello, R., Carpentieri, M. and Kläui, M., 2016. Magnetic skyrmions: from fundamental to applications. *Journal of Physics D: Applied Physics*, *49*(42), p.423001.

4. Bauer, A. and Pfleiderer, C., 2016. Generic aspects of skyrmion lattices in chiral magnets. In *Topological Structures in Ferroic Materials* (pp. 1-28). Springer International Publishing.

5. Nagaosa, N. and Tokura, Y., 2013. Topological properties and dynamics of magnetic skyrmions. *Nature nanotechnology*, *8*(12), pp.899-911.

6. Adams, T., Chacon, A., Wagner, M., Bauer, A., Brandl, G., Pedersen, B., Berger, H., Lemmens, P. and Pfleiderer, C., 2012. Long-wavelength helimagnetic order and skyrmion lattice phase in $Cu_2OSeO_3$. *Physical review letters*, *108*(23), p.237204.

7. Kézsmárki, I., Bordács, S., Milde, P., Neuber, E., Eng, L.M., White, J.S., Rønnow, H.M., Dewhurst, C.D., Mochizuki, M., Yanai, K. and Nakamura, H., 2015. Néel-type skyrmion lattice with confined orientation in the polar magnetic semiconductor $GaV_4S_8$. *Nature materials*, *14*(11), pp.1116-1122.

8. Tokunaga, Y., Yu, X.Z., White, J.S., Rønnow, H.M., Morikawa, D., Taguchi, Y. and Tokura, Y., 2015. A new class of chiral materials hosting magnetic skyrmions beyond room temperature. *Nature communications*, *6*.

9. Tanigaki, T., Shibata, K., Kanazawa, N., Yu, X., Onose, Y., Park, H.S., Shindo, D. and Tokura, Y., 2015. Real-space observation of short-period cubic lattice of skyrmions in MnGe. *Nano letters*, *15*(8), pp.5438-5442.

10. Bauer, A., Neubauer, A., Franz, C., Münzer, W., Garst, M. and Pfleiderer, C., 2010. Quantum phase transitions in single-crystal $Mn_{1-x}Fe_xSi$ and $Mn_{1-x}Co_xSi$: Crystal growth, magnetization, ac susceptibility, and specific heat. *Physical Review B*, *82*(6), p.064404.

11. Yu, X.Z., Kanazawa, N., Onose, Y., Kimoto, K., Zhang, W.Z., Ishiwata, S., Matsui, Y. and Tokura, Y., 2011. Near room-temperature formation of a skyrmion crystal in thin-films of the helimagnet FeGe. *Nature materials*, *10*(2), pp.106-109.

12. Koretsune, T., Nagaosa, N. and Arita, R., 2015. Control of Dzyaloshinskii-Moriya interaction in $Mn_{1-x}Fe_xGe$: a first-principles study. *Scientific reports*, *5*.

13. Gayles, J., Freimuth, F., Schena, T., Lani, G., Mavropoulos, P., Duine, R.A., Blügel, S., Sinova, J. and Mokrousov, Y., 2015. Dzyaloshinskii-Moriya Interaction and Hall Effects in the Skyrmion Phase of $Mn_{1-x}Fe_xGe$. *Physical review letters*, *115*(3), p.036602.

14. Díaz, S.A. and Troncoso, R.E., 2016. Controlling skyrmion helicity via engineered Dzyaloshinskii–Moriya interactions. *Journal of Physics: Condensed Matter*, *28*(42), p.426005.

15. Banerjee, S., Rowland, J., Erten, O. and Randeria, M., 2014. Enhanced stability of skyrmions in two-dimensional chiral magnets with Rashba spin-orbit coupling. *Physical Review X*, *4*(3), p.031045.

16. Grigoriev, S.V., Chernyshov, D., Dyadkin, V.A., Dmitriev, V., Moskvin, E.V., Lamago, D., Wolf, T., Menzel, D., Schoenes, J., Maleyev, S.V. and Eckerlebe, H., 2010. Interplay between crystalline chirality and magnetic structure in $Mn_{1-x}Fe_xSi$. *Physical Review B*, *81*(1), p.012408.

17. Bak, P. and Jensen, M.H., 1980. Theory of helical magnetic structures and phase transitions in MnSi and FeGe. *Journal of Physics C: Solid State Physics*, *13*(31), p.L881.

18. Nakanishi, O., Yanase, A., Hasegawa, A. and Kataoka, M., 1980. The origin of the helical spin density wave in MnSi. *Solid State Communications*, *35*(12), pp.995-998.

19. Maleyev, S.V., 2006. Cubic magnets with Dzyaloshinskii-Moriya interaction at low temperature. *Physical Review B*, *73*(17), p.174402

20. Pfleiderer, C., Julian, S.R. and Lonzarich, G.G., 2001. Non-Fermi-liquid nature of the normal state of itinerant-electron ferromagnets. *Nature*,*414*(6862), pp.427-430.

21. Manyala, N., Sidis, Y., DiTusa, J.F., Aeppli, G., Young, D.P. and Fisk, Z., 2004. Large anomalous Hall effect in a silicon-based magnetic semiconductor. *Nature materials*, *3*(4), pp.255-262.

22. Manyala, N., DiTusa, J.F., Aeppli, G. and Ramirez, A.P., 2008. Doping a semiconductor to create an unconventional metal. *Nature*, *454*(7207), pp.976-980.

23. Panfilov, A.S., 1999. Effect of pressure on magnetic properties of the compound MnSi. *Low Temperature Physics*, *25*, pp.432-435.

24. Chacon, A., Bauer, A., Adams, T., Rucker, F., Brandl, G., Georgii, R., Garst, M. and Pfleiderer, C., 2015. Uniaxial pressure dependence of magnetic order in MnSi. *Physical review letters*, *115*(26), p.267202.

25. Münzer, W., Neubauer, A., Adams, T., Mühlbauer, S., Franz, C., Jonietz, F., Georgii, R., Böni, P., Pedersen, B., Schmidt, M. and Rosch, A., 2010. Skyrmion lattice in the doped semiconductor $Fe_{1-x}Co_xSi$. *Physical Review B*, *81*(4), p.041203.

26. Grigoriev, S.V., Potapova, N.M., Siegfried, S.A., Dyadkin, V.A., Moskvin, E.V., Dmitriev, V., Menzel, D., Dewhurst, C.D., Chernyshov, D., Sadykov, R.A. and Fomicheva, L.N., 2013. Chiral Properties of Structure and Magnetism in $Mn_{1-x}Fe_xGe$ Compounds: When the Left and the Right are Fighting, Who Wins? *Physical review letters*, *110*(20), p.207201.

27. Grigoriev, S.V., Dyadkin, V.A., Moskvin, E.V., Lamago, D., Wolf, T., Eckerlebe, H. and Maleyev, S.V., 2009. Helical spin structure of $Mn_{1-y}Fe_ySi$ under a magnetic field: Small angle neutron diffraction study. *Physical Review B*, *79*(14), p.144417.

28. Grigoriev, S.V., Dyadkin, V.A., Maleyev, S.V., Menzel, D., Schoenes, J., Lamago, D., Moskvin, E.V. and Eckerlebe, H., 2010. Noncentrosymmetric cubic helical ferromagnets Mn$_{1-y}$Fe$_y$Si and Fe$_{1-x}$Co$_x$Si. *Physics of the Solid State*, *52*(5), pp.907-913

29. Cheong, S.W. and Mostovoy, M., 2007. Multiferroics: a magnetic twist for ferroelectricity. *Nature materials*, *6*(1), pp.13-20.

30. Dmitrienko, V.E., Ovchinnikova, E.N., Collins, S.P., Nisbet, G., Beutier, G., Kvashnin, Y.O., Mazurenko, V.V., Lichtenstein, A.I. and Katsnelson, M.I., 2014. Measuring the Dzyaloshinskii-Moriya interaction in a weak ferromagnet. *Nature Physics*, *10*(3), pp.202-206.

31. Shibata, K., Yu, X.Z., Hara, T., Morikawa, D., Kanazawa, N., Kimoto, K., Ishiwata, S., Matsui, Y. and Tokura, Y., 2013. Towards control of the size and helicity of skyrmions in helimagnetic alloys by spin-orbit coupling. *Nature nanotechnology*, *8*(10), pp.723-728.

32. Kanazawa, N., Shibata, K. and Tokura, Y., 2016. Variation of spin–orbit coupling and related properties in skyrmionic system Mn$_{1-x}$Fe$_x$Ge. *New Journal of Physics*, *18*(4), p.045006.

33. See supplementary materials for details of sample preparation and characterizations, ac susceptibility measurements and small angle neutron scattering measurements. The supplemental section also presents the variation of $T_C$ and $AM_S/a^2$ for a wide range of *B20* materials.

34. Rhodes, P. and Wohlfarth, E.P., 1963, May. The effective Curie-Weiss constant of ferromagnetic metals and alloys. In *Proceedings of the Royal Society of London A: Mathematical, Physical and Engineering Sciences* (Vol. 273, No. 1353, pp. 247-258). The Royal Society.

35. Moriya, T. and Takahashi, Y., 1984. Itinerant electron magnetism. *Annual Review of Materials Science*, *14*(1), pp.1-25.

36. Dhital, C., Khan, M.A., Saghayezhian, M., Phelan, W.A., Young, D.P., Jin, R.Y. and DiTusa, J.F., 2017. Effect of negative chemical pressure on the prototypical itinerant magnet MnSi. *Physical Review B*, *95*(2), p.024407.

37. Grigoriev, S.V., Maleyev, S.V., Okorokov, A.I., Chetverikov, Y.O., Böni, P., Georgii, R., Lamago, D., Eckerlebe, H. and Pranzas, K., 2006. Magnetic structure of MnSi under an applied field probed by polarized small-angle neutron scattering. *Physical Review B*, *74*(21), p.214414.

38. Grigoriev, S.V., Maleyev, S.V., Okorokov, A.I., Chetverikov, Y.O. and Eckerlebe, H., 2006. Field-induced reorientation of the spin helix in MnSi near T c. *Physical Review B*, *73*(22), p.224440.

39. Grigoriev, S.V., Maleyev, S.V., Okorokov, A.I., Chetverikov, Y.O., Georgii, R., Böni, P., Lamago, D., Eckerlebe, H. and Pranzas, K., 2005. Critical fluctuations in MnSi near $T_C$: A polarized neutron scattering study. *Physical Review B*, *72*(13), p.134420.

40. Grigoriev, S.V., Maleyev, S.V., Moskvin, E.V., Dyadkin, V.A., Fouquet, P. and Eckerlebe, H., 2010. Crossover behavior of critical helix fluctuations in MnSi. *Physical Review B*, *81*(14), p.144413.

41. Levatić, I., Popčević, P., Šurija, V., Kruchkov, A., Berger, H., Magrez, A., White, J.S., Rønnow, H.M. and Živković, I., 2016. Dramatic pressure-driven enhancement of bulk skyrmion stability. *Scientific reports*, *6*.

42. Huang, S.X. and Chien, C.L., 2012. Extended skyrmion phase in epitaxial FeGe (111) thin films. *Physical Review Letters*, *108*(26), p.267201.

43. Harris, P., Lebech, B., Shim, H.S., Mortensen, K. and Pedersen, J.S., 1995. Small-angle neutron-scattering studies of the magnetic phase diagram of MnSi. *Physica B: Condensed Matter*, *213*, pp.375-377.

44. Thessieu, C., Flouquet, J., Lapertot, G., Stepanov, A.N. and Jaccard, D., 1995. Magnetism and spin fluctuations in a weak itinerant ferromagnet: MnSi. *Solid state communications*, *95*(10), pp.707-712.

45. Belitz, D., Kirkpatrick, T.R. and Rollbühler, J., 2005. Tricritical behavior in itinerant quantum ferromagnets. *Physical Review Letters*, *94*(24), p.247205.

46. Zhang, L., Menzel, D., Jin, C., Du, H., Ge, M., Zhang, C., Pi, L., Tian, M. and Zhang, Y., 2015. Critical behavior of the single-crystal helimagnet MnSi. *Physical Review B*, *91*(2), p.024403.

47. Bauer, A., Garst, M. and Pfleiderer, C., 2013. Specific heat of the skyrmion lattice phase and field-induced tricritical point in MnSi. *Physical Review Letters*, *110*(17), p.177207.

48. Pappas, C., Bannenberg, L.J., Lelièvre-Berna, E., Qian, F., Dewhurst, C.D., Dalgliesh, R.M., Schlagel, D.L., Lograsso, T.A. and Falus, P., 2017. Magnetic Fluctuations, Precursor Phenomena, and Phase Transition in MnSi under a Magnetic Field. *Physical Review Letters*, *119*(4), p.047203.

49. Teyssier, J., Giannini, E., Guritanu, V., Viennois, R., Van Der Marel, D., Amato, A. and Gvasaliya, S.N., 2010. Spin-glass ground state in $Mn_{1-x}Co_xSi$. *Physical Review B*, *82*(6), p.064417.

50. Beille, J., Voiron, J. and Roth, M., 1983. Long period helimagnetism in the cubic B20 $Fe_xCo_{1-x}Si$ and $Co_xMn_{1-x}Si$ alloys. *Solid state communications*, *47*(5), pp.399-402.

51. Ishikawa, Y., Shirane, G., Tarvin, J.A. and Kohgi, M., 1977. Magnetic excitations in the weak itinerant ferromagnet MnSi. *Physical Review B*, *16*(11), p.4956.

52. Blundell, S., Magnetism in condensed matter, Oxford Master Series in Physics, Oxford University Press (New York) (2001).

53. DiTusa, J.F., Zhang, S.B., Yamaura, K., Xiong, Y., Prestigiacomo, J.C., Fulfer, B.W., Adams, P.W., Brickson, M.I., Browne, D.A., Capan, C. and Fisk, Z., 2014. Magnetic, thermodynamic, and electrical transport properties of the noncentrosymmetric *B20* germanides MnGe and CoGe. *Physical Review B*, *90*(14), p.144404

54. Martin, N., Deutsch, M., Chaboussant, G., Damay, F., Bonville, P., Fomicheva, L.N., Tsvyashchenko, A.V., Rössler, U.K. and Mirebeau, I., 2017. Long-period helical structures and twist-grain boundary phases induced by chemical substitution in the $Mn_{1-x}(Co, Rh)_xGe$ chiral magnet. *Physical Review B*, *96*(2), p.020413.

55. Grigoriev, S.V., Siegfried, S.A., Altynbayev, E.V., Potapova, N.M., Dyadkin, V., Moskvin, E.V., Menzel, D., Heinemann, A., Axenov, S.N., Fomicheva, L.N. and Tsvyashchenko, A.V., 2014. Flip of spin helix chirality and ferromagnetic state in $Fe_{1-x}Co_xGe$ compounds. *Physical Review B* **90**, p.174414.

56. Onose, Y., Takeshita, N., Terakura, C., Takagi, H. and Tokura, Y., 2005. Doping dependence of transport properties in $Fe_{1-x}Co_xSi$. *Physical Review B* **72**, p.224431.

# Supplementary materials

**Exploring the origins of the Dzyalloshinskii-Moriya interaction in MnSi**

C. Dhital[1]*, L. DeBeer-Schmitt[2], Q. Zhang[1, 2], W. Xie[3], D. P. Young[1], J. F. DiTusa[1]**

*1 Department of Physics and Astronomy, Louisiana State University, Baton Rouge, LA 70803*

*2 Oak Ridge National Laboratory, Oak Ridge, Tennessee 37831, USA*

*3 Department of Chemistry, Louisiana State University, Baton Rouge, LA, 70803, USA*

In this supplemental materials document we provide details about the sample synthesis and characterization. In addition, we provide a more complete presentation of our ac susceptibility data and outline our procedure for analyzing the small angle neutron scattering data. We also summarize our procedure and provide data used to determine the universality class of the phase transition, as well as details of our comparison of the spin wave stiffness and the critical temperature of our samples.

## 1. Sample preparation and Characterization:

The samples were prepared by loading arc melted polycrystalline pellets into a graphite crucible and using a modified Bridgman method under a flowing argon environment. The phase purity and crystallinity were checked using powder and single crystal X-ray diffraction, whereas the stoichiometry was obtained using single crystal X-ray refinement and Wavelength Dispersive Spectroscopy (WDS). The x-ray refinement and WDS methods give consistent values of chemical occupancies. We find that there is a progressive increase of lattice parameter, *a*, with iridium concentration, *x* (Fig. S1), indicating that Ir successfully replaces Mn in our samples.

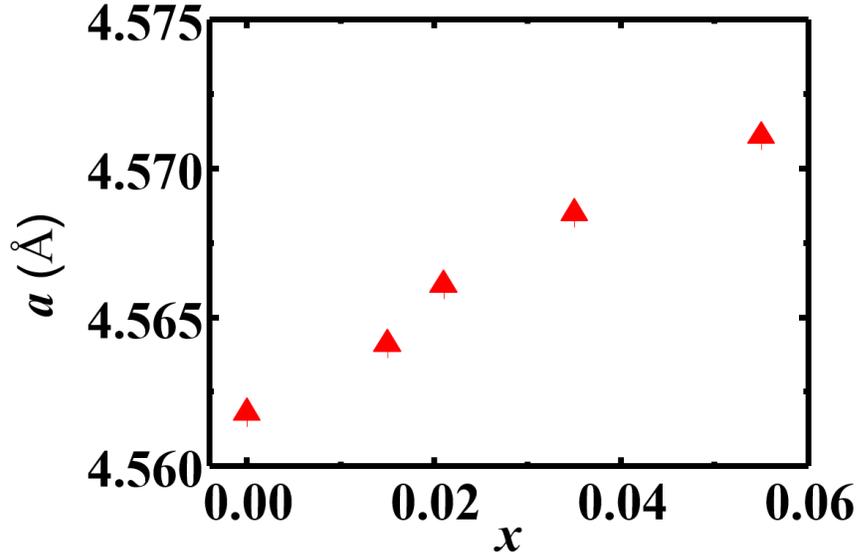

**Fig. S1** Variation of lattice parameter *a* as a function of *x* in Mn$_{1-x}$Ir$_x$Si.

## 2. Magnetic measurements

Magnetization measurements, both ac and dc, were carried out in a Quantum Design 7-T MPMS SQUID magnetometer. The results are presented in Fig. 1 of the main text. All the samples exhibit a Curie-Weiss behavior in their paramagnetic regime. The results of fitting a modified Curie-Weiss form, $\chi = \chi_0 + \frac{C}{T-\theta}$, to the data are presented in Table S1. The ac susceptibility as a function of applied magnetic field, *H*, was measured at a frequency of 100 Hz and an amplitude of 1 Oe in a Quantum Design MPMS-7 for all crystals. The results for temperatures below the Curie point are presented in Fig. S2. The values of the critical fields $H_{C1}$, $H_{A1}$, $H_{A2}$, and $H_{C2}$ are obtained from these data as described in the main text.

**Table S1: Parameters resulting from fits of dc susceptibility data by a modified Curie-Weiss fit for Mn$_{1-x}$Ir$_x$Si**

| *Concentration x* | *Susceptibility* $\chi_0$ *(emu/mol-Oe)* | *Curie constant* *C (emu K/mol-Oe)* | *Curie-Weiss temperature* $\theta$ *(K)* |
|---|---|---|---|
| 0 | (5.0±0.2)x10$^{-5}$ | 0.59±0.01 | 30.5±0.1 |
| 0.015 | (1.7±0.1)x10$^{-4}$ | 0.57±0.01 | 25.1±0.1 |
| 0.021 | (3.2±0.1)x10$^{-4}$ | 0.56±0.01 | 20.5±0.1 |
| 0.035 | (2.7±0.1)x10$^{-4}$ | 0.49±0.01 | 18.7±0.1 |
| 0.055 | (2.0±0.1)x10$^{-4}$ | 0.45±0.01 | 8.1±0.1 |

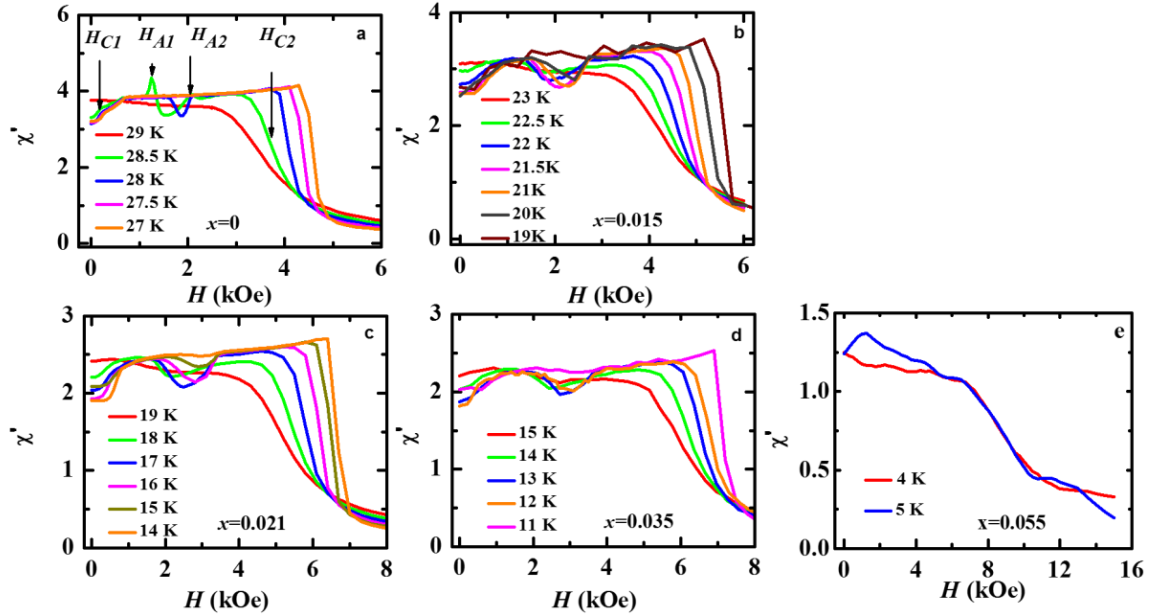

**Fig. S2** ac susceptibility. The real part of the ac susceptibility, $\chi'$, as a function of dc magnetic field for (a) $x=0$ ($T_C=29$ K), (b) $x=0.015$ ($T_C=23$ K), (c) $x=0.021$ ($T_C=19$ K), (d) $x=0.035$ ($T_C=15$ K), and (e) $x=0.055$ ($T_C=6$ K). The critical field values $H_{C1}$, $H_{A1}$, $H_{A2}$, and $H_{C2}$ for the $x=0$ sample are indicated by the arrows in frame a.

## 2. Determination of helix wave vector *k* and the integrated intensity *I*

Small Angle Neutron Scattering (SANS) measurements were carried out in the GP-SANS beamline at the High Flux Isotope Reactor (HFIR) at Oak Ridge National Laboratory (ORNL). The mean wavelength of the incident neutrons employed was $\lambda = 4.75$ Å with $\Delta\lambda/\lambda = 0.16$. The sample-to-detector distance was 8.65 m. The sample was aligned such that incident beam ($K_i$) and magnetic field ($H$) both are parallel to the [1 -1 0] direction and perpendicular to the detector. The [1 1 1] direction was aligned so that it was nearly along the horizontal direction. In this configuration, we expect 2 pairs of peaks corresponding to the [111] and [11-1] directions in the helical phase (Fig. S3a). The acute angle between these two directions is 70.5°. The samples used in this study were large (~500 mg) and typically consisted of several crystal domains. The existence of several crystalline domains can result in an artificial broadening of the peaks in the azimuthal direction due to contamination from possibly misaligned magnetic domains. However, for every sample measured, we observe at least two pairs of peaks (at zero field) that are 70.5°, indicating that a large crystal domain was well aligned in the manner intended. We have carried out our analysis using a single pair of peaks that originates from a single crystalline domain. To ensure that we recover the full peak intensity for every temperature and field measured, we performed a rocking scan of 20 degrees (-10 to 10 degrees from the central peak position), summing the intensity over the entire scan. To determine $k$ we fit a Gaussian curve to the intensity as shown in Fig. S3 (b). The center position of each peak corresponds to the helix wave vector $k$.

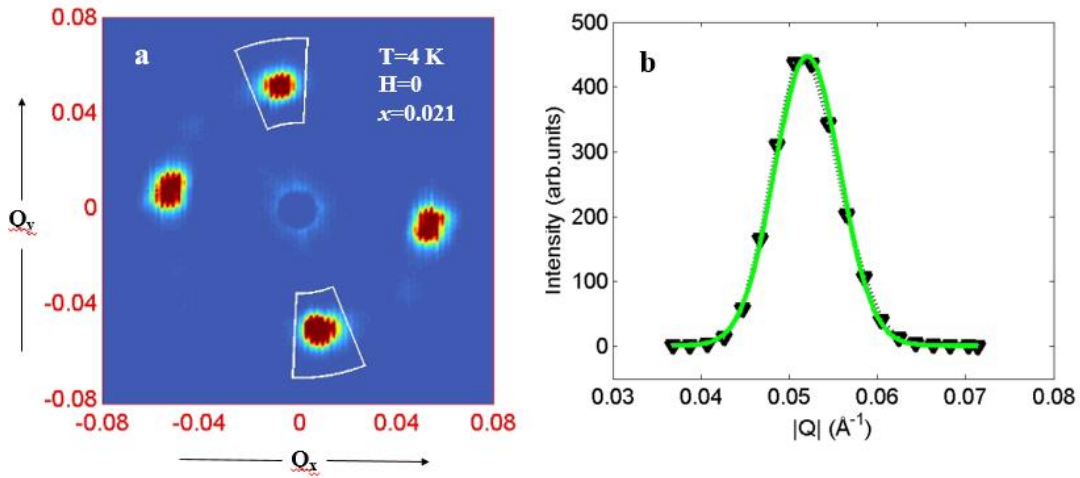

**Fig S3**: Small angle neutron scattering. (a) Typical scattering pattern for the helical phase. (b) The combined integrated intensity of peaks enclosed by the white boxes in frame (a) along with a Gaussian fit to the peak (green line).

The integrated intensity, $I$, was obtained from fits of a Gaussian form to the data. To characterize the magnetic transition, the temperature dependence of the integrated intensity was fit by the power law of the form $I=I_0(1-T/T_C)^{2\beta}$, as shown in Fig S4. The best-fit values for the parameters $\beta$ and $T_C$ are indicated in the figure.

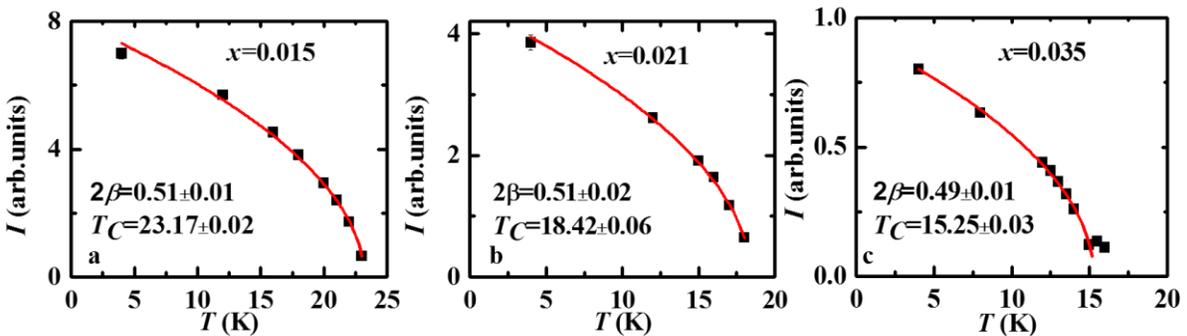

**Fig S4**: Power law fit of the form $I = I_0 (1-T/T_C)^{2\beta}$ to the integrated intensity of the peaks associated with the helical magnetic state enclosed by the white boxes in Fig. S3a for $Mn_{1-x}Ir_xSi$.

The SANS intensity patterns taken just above $T_C$ (Fig. S5) display rings of scattering, indicating the presence of a fluctuating precursor phase consistent with previous studies on MnSi [1,2].

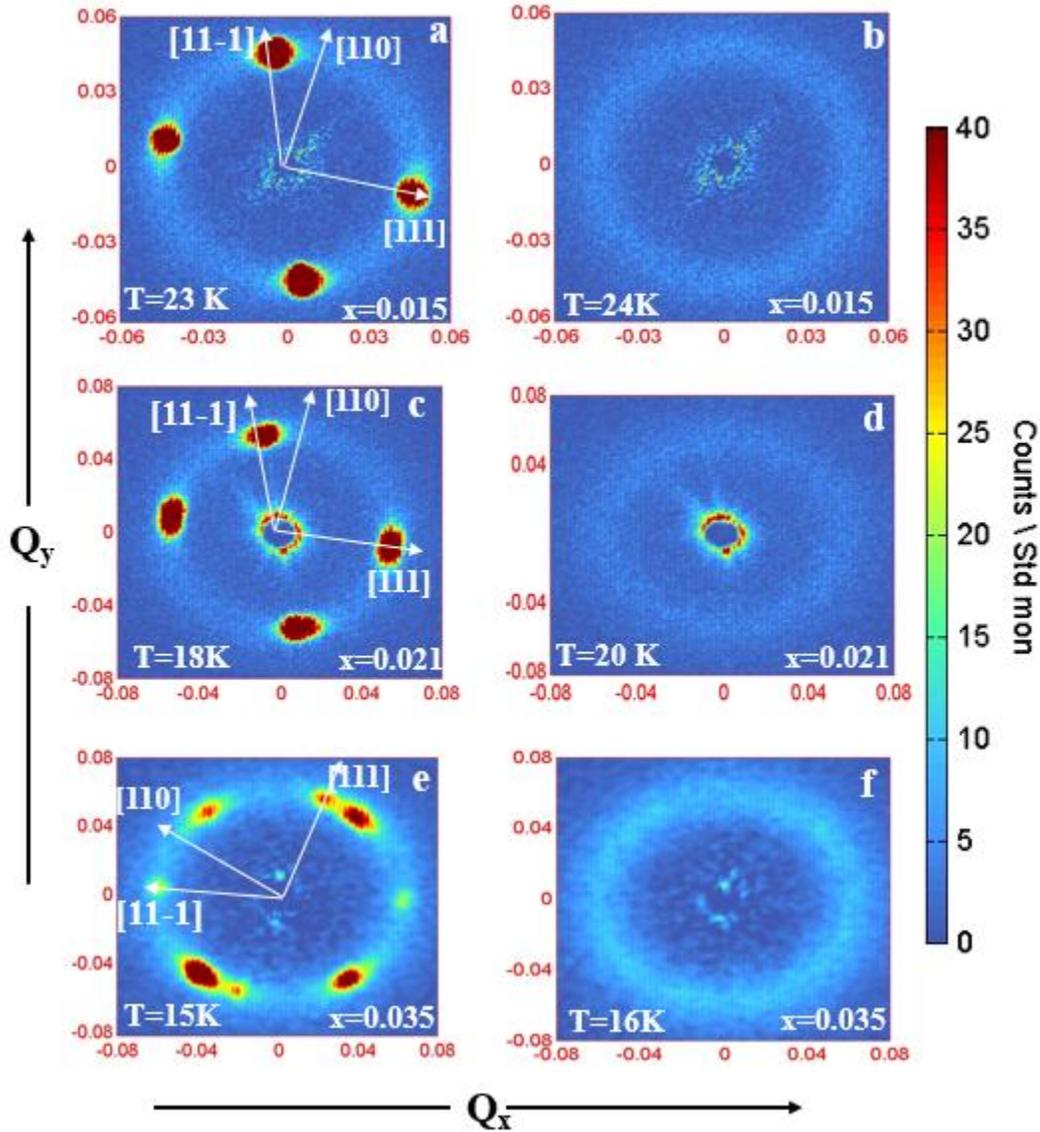

Fig S5: SANS intensity pattern at zero field taken just below (a, c, and e) and just above (b, d, and f) the magnetic critical temperature for $Mn_{1-x}Ir_xSi$ samples (temperatures and substitution levels are indicted in the figure).

We have also performed temperature scans at finite fields for temperatures close to $T_C$ to verify the presence of the skyrmion lattice or A-phase. The variation of integrated intensities in a region around [110] as a function of temperature and field for all 3 samples is presented in Fig. S6.

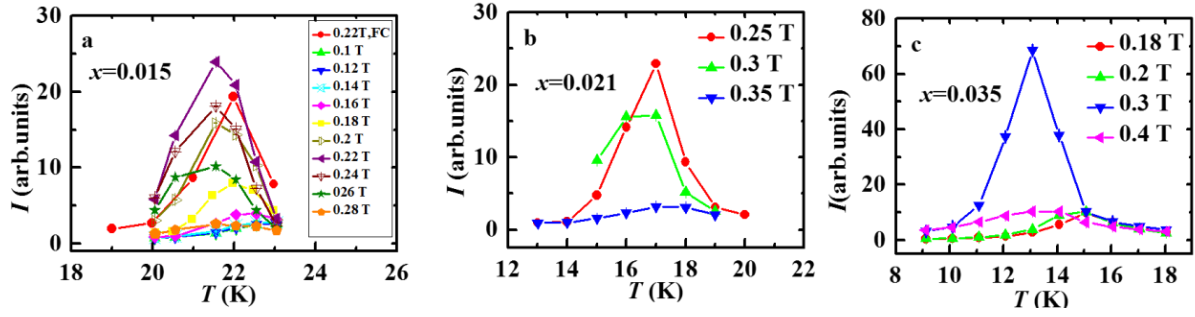

Fig S6: Variation of the integrated intensity in the region around the [110] peak position as a function of field and temperature for $Mn_{1-x}Ir_xSi$ samples. Strong intensities correspond to a well-established A-phase.

## 3. Comparison of the Curie temperature to estimates of the spinwave stiffness parameter for the cubic *B20* structured materials.

In the mean field approximation $T_C \propto JS^2$ [2], so that use of the relations $SD=kA$ and $A \propto JSa^{2}$, yields the form $T_C \propto AS/a^2$. The variation of $T_C$ with $AM_S/a^2$ for a large number of *B20* materials spanning a wide range of Curie temperatures is presented in Fig. S7. Here we have assumed that $S=M_S$ and have made use of the approximation that $g\mu_B H_{C2}=Ak^2$. It is clear that a simple proportionality of $T_C$ to $AM_S/a^2$ does not describe these data well. We have included the results of fitting of a linear dependence and a power law form $T_C \propto \left(\frac{AM_S}{a^2}\right)^r$ to the MnSi derived materials for comparison to this wider set of data.

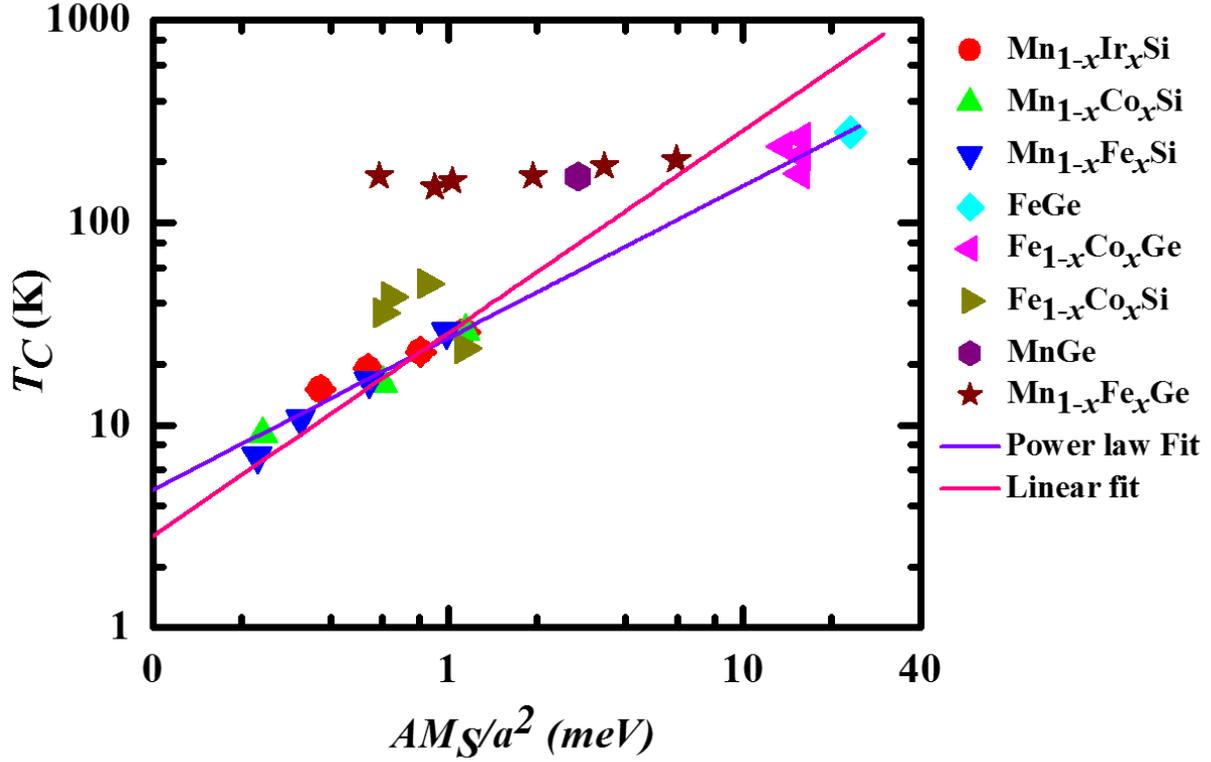

**Fig S7**: Variation of $T_C$ as a function of $AM_S/a^2$ for *B20* materials. Lines represent the best fit of a power law (violet) $T_C \propto \left(\frac{AM_S}{a^2}\right)^r$ and linear (pink) $T_C \propto \left(\frac{AM_S}{a^2}\right)$ to the data for the Mn-based silicides with $r=(0.75\pm0.03)$. The data for MnGe, $Mn_{1-x}Fe_xGe$, $Fe_{1-x}Co_xGe$, $Fe_{1-x}Co_xSi$, $Mn_{1-x}Co_xSi$, $Mn_{1-x}Fe_xSi$ are reproduced from data in Refs. [6-14] with permission from the publishers and/or authors.

# References for supplementary materials

1. Grigoriev, S.V., Maleyev, S.V., Okorokov, A.I., Chetverikov, Y.O., Georgii, R., Böni, P., Lamago, D., Eckerlebe, H. and Pranzas, K., 2005. Critical fluctuations in MnSi near $T_C$: A polarized neutron scattering study. *Physical Review B 72*(13), p.134420.
2. Blundell, S., Magnetism in condensed matter, Oxford Master Series in Physics, Oxford University Press (New York) (2001).
3. Grigoriev, S.V., Maleyev, S.V., Moskvin, E.V., Dyadkin, V.A., Fouquet, P. and Eckerlebe, H., 2010. Crossover behavior of critical helix fluctuations in MnSi. *Physical Review B 81*, p.144413.
4. Bak, P. and Jensen, M.H., 1980. Theory of helical magnetic structures and phase transitions in MnSi and FeGe. *Journal of Physics C: Solid State Physics 13*, p.L881.

5. Maleyev, S.V., 2006. Cubic magnets with Dzyaloshinskii-Moriya interaction at low temperature. *Physical Review B* **73**, p.174402
6. DiTusa, J.F., Zhang, S.B., Yamaura, K., Xiong, Y., Prestigiacomo, J.C., Fulfer, B.W., Adams, P.W., Brickson, M.I., Browne, D.A., Capan, C. and Fisk, Z., 2014. Magnetic, thermodynamic, and electrical transport properties of the noncentrosymmetric *B20* germanides MnGe and CoGe. *Physical Review B* **90**, p.144404
7. Martin, N., Deutsch, M., Chaboussant, G., Damay, F., Bonville, P., Fomicheva, L.N., Tsvyashchenko, A.V., Rössler, U.K. and Mirebeau, I., 2017. Long-period helical structures and twist-grain boundary phases induced by chemical substitution in the $Mn_{1−x}(Co, Rh)_xGe$ chiral magnet. *Physical Review B* **96**, p.020413.
8. Bauer, A., Neubauer, A., Franz, C., Münzer, W., Garst, M. and Pfleiderer, C., 2010. Quantum phase transitions in single-crystal $Mn_{1−x}Fe_xSi$ and $Mn_{1−x}Co_xSi$: Crystal growth, magnetization, ac susceptibility, and specific heat. *Physical Review B* **82**, p.064404.
9. Grigoriev, S.V., Dyadkin, V.A., Moskvin, E.V., Lamago, D., Wolf, T., Eckerlebe, H. and Maleyev, S.V., 2009. Helical spin structure of $Mn_{1−y}Fe_ySi$ under a magnetic field: Small angle neutron diffraction study. *Physical Review B* **79**, p.144417.
10. Teyssier, J., Giannini, E., Guritanu, V., Viennois, R., Van Der Marel, D., Amato, A. and Gvasaliya, S.N., 2010. Spin-glass ground state in $Mn_{1−x}Co_xSi$. *Physical Review B* **82**, p.064417.
11. Beille, J., Voiron, J. and Roth, M., 1983. Long period helimagnetism in the cubic B20 $Fe_xCo_{1−x}Si$ and $Co_xMn_{1−x}Si$ alloys. *Solid State Communications*, **47**, pp.399-402.
12. Grigoriev, S.V., Siegfried, S.A., Altynbayev, E.V., Potapova, N.M., Dyadkin, V., Moskvin, E.V., Menzel, D., Heinemann, A., Axenov, S.N., Fomicheva, L.N. and Tsvyashchenko, A.V., 2014. Flip of spin helix chirality and ferromagnetic state in $Fe_{1−x}Co_xGe$ compounds. *Physical Review B* **90**, p.174414.
13. Onose, Y., Takeshita, N., Terakura, C., Takagi, H. and Tokura, Y., 2005. Doping dependence of transport properties in $Fe_{1−x}Co_xSi$. *Physical Review B* **72**, p.224431.
14. Kanazawa, N., Shibata, K. and Tokura, Y., 2016. Variation of spin–orbit coupling and related properties in skyrmionic system $Mn_{1-x}Fe_xGe$. *New Journal of Physics* **18**, p.045006.